\documentclass[10pt,conference]{IEEEtran}
\usepackage[utf8]{inputenc}
\usepackage[noadjust]{cite}

\usepackage{graphicx}
\usepackage{amsmath}
\usepackage{amssymb}
\usepackage{mathtools}

\usepackage{caption}
\usepackage{subcaption}

\usepackage{tikz}
\usepackage{pgfplots}
\usepgfplotslibrary{units}
\usetikzlibrary{arrows}
\usetikzlibrary{automata,positioning}

\definecolor{Set1-7-1}{RGB}{228,26,28}
\definecolor{Set1-7-2}{RGB}{55,126,184}
\definecolor{Set1-7-3}{RGB}{77,175,74}
\definecolor{Set1-7-4}{RGB}{152,78,163}
\definecolor{Set1-7-5}{RGB}{255,127,0}
\definecolor{Set1-7-6}{RGB}{166,86,40}
\definecolor{Set1-7-7}{RGB}{0,0,0}

\DeclareMathOperator*{\argm}{arg}

\newcommand{\ctau}{\lceil \tau \rceil}
\newcommand{\supk}{^{(k)}}
\newcommand{\supkM}{^{(k-1)}}

\begin{document}

\author{\IEEEauthorblockN{Mathieu Xhonneux,\IEEEauthorrefmark{2} Joachim Tapparel,\IEEEauthorrefmark{1} Orion Afisiadis,\IEEEauthorrefmark{1} Alexios Balatsoukas-Stimming,\IEEEauthorrefmark{3}} and Andreas Burg\IEEEauthorrefmark{1}\\%
	\IEEEauthorblockA{\IEEEauthorrefmark{1}Telecommunication Circuits Laboratory, EPFL, Switzerland\\
	\IEEEauthorrefmark{2}ICTEAM, UCLouvain, Belgium\\
		\IEEEauthorrefmark{3}Department of Electrical Engineering, Eindhoven University of Technology, The Netherlands}
}

\title{A Maximum-Likelihood-based Multi-User LoRa Receiver Implemented in GNU Radio}
\maketitle

\medskip

\begin{abstract} 
    LoRa is a popular low-power wide-area network
    (LPWAN) technology that uses spread-spectrum to achieve long-
    range connectivity and resilience to noise and interference. For
    energy efficiency reasons, LoRa adopts a pure ALOHA access
    scheme, which leads to reduced network throughput due to
    packet collisions at the gateways. To alleviate this issue, in this
    paper we analyze and implement a LoRa receiver that is able to
    decode LoRa packets from two interfering users. Our main contribution
    is a two-user detector derived in a maximum-likelihood
    fashion using a detailed interference model. As the complexity
    of the maximum-likelihood sequence estimation is prohibitive, a
    complexity-reduction technique is introduced to enable a practical
    implementation of the proposed two-user detector. This detector
    has been implemented along with an interference-robust synchronization 
	algorithm on the GNU Radio software-defined radio (SDR)
    platform. The SDR implementation shows the effectiveness of the
    proposed method and also allows its experimental evaluation.
    Measurements indicate that our detector inherently leverages
    the time offset between the two colliding users to separate and
    demodulate their signals.
\end{abstract} 

\section{Introduction}

With the rise of the Internet of Things (IoT), new low-power wide-area network (LPWAN) standards have emerged~\cite{raza2017low}.
Unlike 3GPP cellular standards such as NB-IoT, LPWAN standards rely on low-complexity PHY and MAC layers.
In the last years, LoRa has become one of the most popular LPWAN protocols~\cite{haxhibeqiri2018survey}.
At the PHY layer, LoRa uses a chirp spread-spectrum modulation that enables receivers to attain sensitivities as low as $-130$~dBm~\cite{augustin2016study}.
The spreading gain of the LoRa modulation is determined by the spreading factor (SF), which allows to trade off transmission time and data rate for coverage.
For energy efficiency reasons, LoRaWAN relies on a non-slotted ALOHA multiple access scheme. As the user end-nodes are not synchronized, interference between users is very common in large-scale LoRa networks, posing
 a threat to the scalability of future massive LoRa systems.

Interference between LoRa users can be one of two very different types: \textit{inter-SF interference} and \textit{same-SF interference}. Due to spreading, the impact of
inter-SF interference is different than the impact of same-SF interference.
In particular, inter-SF interference can be treated as additional AWGN~\cite{croce2019lora} that slightly degrades the signal-to-noise ratio (SNR). On the other hand, interfering transmissions which use the same SF lead to strong interference at the receiver as they are far from orthogonal. Nevertheless, colliding LoRa packets with the same SF exhibit a capture effect, such that the strongest signal can often be demodulated at the expense of the weakest signal~\cite{afisiadis2019error} if the signal-to-interference ratio (SIR) and the SNR are sufficient. 
Furthermore, experimental measurements have shown that for a 3~dB difference in power between two users, there is a 97\% chance of demodulating the packet of the strongest user \cite{fernandes2019real}, but the packet of the weaker user is generally lost.

To improve the throughput of LoRa networks, several LoRa multi-user receivers have been suggested recently \cite{temim2020enhanced,tong2020combating,hu2020sclora,xia2020ftrack}.
In \cite{temim2020enhanced}, a successive interference cancellation (SIC) receiver that relies on the conventional single-user detector is proposed. 
A multi-user receiver based on non-stationary signal scaling to separate superimposed transmissions is presented in \cite{tong2020combating}.
In \cite{hu2020sclora} and \cite{xia2020ftrack}, the authors suggest to first demodulate all symbols and then to assign them to their respective user based on transmitter-specific features (e.g., time offset and power).
The aforementioned works have established the challenge of designing robust multi-user LoRa receivers that will allow LoRaWAN to scale by exploiting the concept of non-orthogonal multiple access (NOMA). However, a multi-user receiver that is derived based on the maximum likelihood (ML) principle, as well as its practical implementation and experimental evaluation, are still lacking in the literature.

\subsubsection*{Contributions} 
In this work, we present a two-user LoRa detector derived from the maximum likelihood (ML) principle.
Due to the high complexity of the ML detector, we also propose complexity reduction techniques that enable a practical implementation of the receiver.
The proposed detector requires estimates of the power, the carrier frequency offset (CFO), and the sampling time offset (STO) of each user.
We therefore also design a synchronization algorithm capable of detecting the arrival of a new user and of estimating its parameters, even in the presence of an on-going transmission with the same SF.
We implement the proposed multi-user detector and the synchronization algorithm with a software-defined radio (SDR). Our GNU Radio implementation of the multi-user LoRa receiver is publicly available at~\cite{2020GNUrepo}.
We also present a performance evaluation of the proposed detector, showing the capability of the receiver to demodulate two overlapping LoRa users with low error rates. Finally, we discuss the impact of integer and fractional sampling time offsets between the colliding users, which have an important impact on the ability to separate the two signals.

\section{Principles of the LoRa PHY}
\label{sec:lora}

In this section, we provide a brief summary of the LoRa modulation, the single-user demodulation stage, and the structure of the preamble.
We then explain the baseband-equivalent model of two superimposed LoRa signals.

\subsection{Modulation and Demodulation}
LoRa is a chirp spread-spectrum modulation operating in the ISM bands, with typical passband bandwidth values $B~\in~\{ 125, 250, 500 \}$~kHz.
LoRa symbols are chirps, i.e., signals whose instantaneous frequency increases linearly and spans the entire bandwidth. Every chirp consists of $N = 2^{\textrm{SF}}$ chips and carries $\textrm{SF}$ bits of information,
where $\textrm{SF} \in \{ 7, \dots , 12\}$ is called the \textit{spreading factor}.
When sampled at the Nyquist frequency $f_s = B$, the discrete-time baseband-equivalent of a symbol is split into $N$ frequency steps \cite{chiani2019lora}, \cite{ghanaatian2019lora}.
For a symbol $s \in \{0, \dots, N-1\}$, the first chip starts at an initial frequency of $\left(\frac{sB}{N} - \frac{N}{2}\right)$.
The instantaneous frequency of the chirp increases by $\frac{B}{N}$ in every chip. When the Nyquist frequency $\frac{B}{2}$ is attained, i.e., at chip index $n_f = N - s$,
a folding to the frequency $-\frac{B}{2}$ occurs.
The corresponding discrete-time baseband-equivalent equation of a LoRa symbol $s$ can be expressed as \cite{chiani2019lora}, \cite{ghanaatian2019lora}
\begin{equation} \label{eq:baseband-full}
    x_{s}[n]=\left\{\begin{array}{ll}
        {e^{j 2 \pi\left(\frac{n^2}{2N} \left( \frac{B}{f_s} \right)^2 + \left(\frac{s}{N}-\frac{1}{2}\right) \left( \frac{B}{f_s} \right) n \right) ,}} & {0 \leq n < n_f,} \\ 
        {e^{j 2 \pi\left(\frac{n^2}{2N} \left( \frac{B}{f_s} \right)^2 + \left(\frac{s}{N}-\frac{3}{2}\right) \left( \frac{B}{f_s} \right) n \right) ,}} & {n_f \leq n < N.}
    \end{array}\right.
\end{equation}

When the transmission takes place over an AWGN channel with a complex-valued channel gain $h \in \mathbb{C}$,
the received LoRa symbol is represented by $y[n] = h x_s[n] + z[n]$,
where $z[n] \sim \mathcal{CN}(0, \sigma^2)$ is complex AWGN with variance $\sigma^2 = \frac{N_0}{2N}$
and $N_0$ is the single-sided noise power spectral density.
A receiver first performs a point-by-point multiplication of the sampled signal $y[n]$ with $x^*_0[n]$, the complex conjugate of an
unmodulated symbol $s = 0$, yielding the \textit{dechirped} signal
\begin{equation} \label{eq:dechirped}
    \tilde{y}[n] = y[n] x^*_0[n] = \sqrt{P} e^{j 2 \pi n \frac{s}{N} + \theta} + \tilde{z}[n],
\end{equation}
where $P = |h|^2$ is the power of the signal at the receiver, $\theta = \angle h$
represents the phase introduced by the channel, and $\tilde{z}[n] = z[n] x^*_0[n]$.
For a perfectly synchronized receiver, $\tilde{y}[n]$ contains a single complex tone of frequency $\frac{s}{N}$ and AWGN.

Let $Y[i] = \sum_{n = 0}^{N-1} \tilde{y}[n] e^{-j 2 \pi \frac{ni}{N}}$ be the $i$-th bin from the $N$-point discrete Fourier transform (DFT) of the dechirped signal.
In the frequency domain, the single complex tone of frequency $\frac{s}{N}$ translates into a Kronecker delta at position $i = s$.
Typical receivers perform non-coherent detection by selecting the DFT bin with the largest magnitude
\begin{equation} \label{eq:su-nc}
    \hat{s} = \argm \max_{s} \left| Y[s] \right|.
\end{equation}

\subsection{LoRa Preamble Structure}

\begin{figure}[t] 
    \centering
    \includegraphics[width=0.49\textwidth]{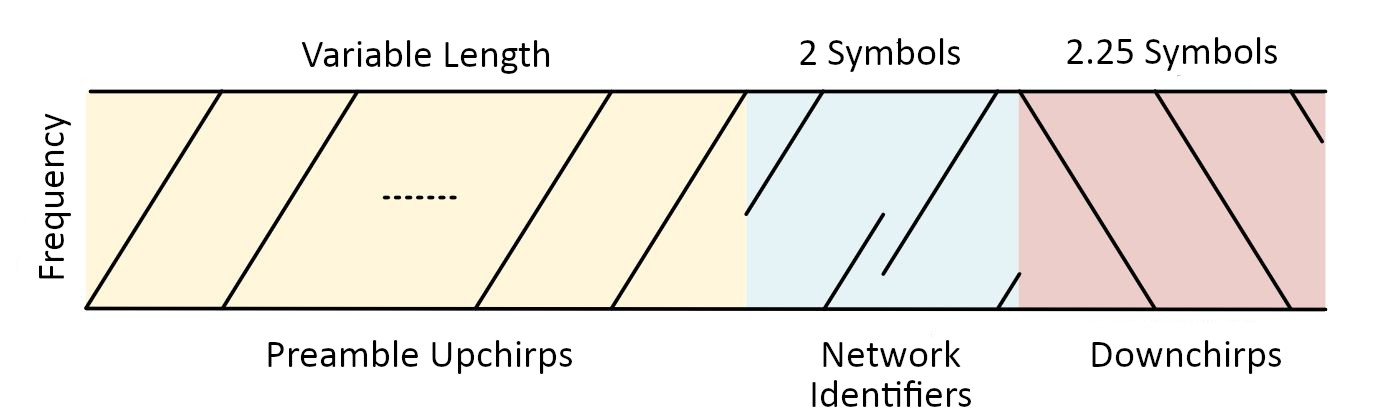}
    \caption{Structure of a LoRa preamble.}
    \label{fig:preamble}
\end{figure}

Every LoRa packet starts with a specific preamble for synchronization purposes.
The preamble consists of $N_{\text{pr}}$ repetitions of an unmodulated upchirp (i.e., $x_0[n]$), succeeded by two symbols acting as network identifiers,
and $2.25$ downchirps $x^*_0[n]$.
The structure of the preamble is illustrated in Fig.~\ref{fig:preamble}.
It is well known that the combination
of upchirps and downchirps can be exploited by a receiver to estimate the CFO and STO of a user~\cite{bernier2020low},~\cite{xhonneux2019low}.
As explained in Section~\ref{sec:receiver}, our two-user receiver also leverages this preamble in its synchronization stage.

\subsection{Signal Model for Two Interfering Users}

\begin{figure}[t] 
    \centering
    \includegraphics[width=0.45\textwidth]{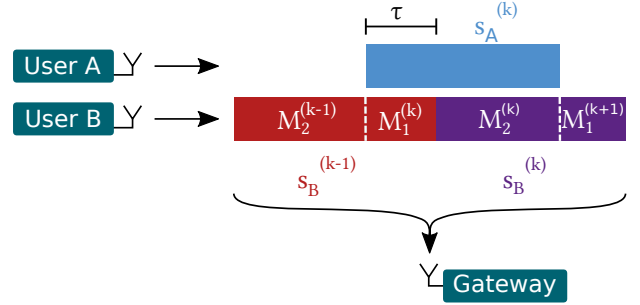}
    \caption{Two-user scenario where user~B has a time offset $\tau$ with respect to user~A.}
    \label{fig:interferences}
\end{figure}

In this section, we describe the model of superimposed signals from two users with the same SF, namely user~A and user~B.
Since LoRa uses a non-slotted ALOHA multiple access scheme, the users are neither synchronized among themselves nor to the gateway.
Let us define $s\supk_{\text{A}}$ and $s\supk_{\text{B}}$ as the $k$-th symbols sent by users~A and~B, respectively.
To simplify the explanation and the mathematical formulation of the multi-user detector, we assume that the gateway is perfectly synchronized in frequency and time to user A, whose packet arrives first. This can be achieved with a standard synchronization procedure~\cite{xhonneux2019low,bernier2020low,tapparel2020open}. We define $\tau \in [0, N)$ as the relative chip-level time offset between the first chip of a symbol transmitted by user A and the first chip of
the next symbol of user B, as illustrated in Fig.~\ref{fig:interferences}.
This offset can be split into an integer part $L_{\textrm{STO}} = \lfloor \tau \rfloor$ and a non-integer part 
$\lambda_{\textrm{STO}} = \tau - \lfloor \tau \rfloor$ \cite{afisiadis2019error}. 

Since the transmission of user B experiences an STO $\tau$ with respect to user A,
the first $\lfloor \tau \rfloor$ samples of symbol $s\supk_{\text{A}}$ overlap with symbol $s\supkM_{\text{B}}$ and the last $N - \lfloor \tau \rfloor$ samples of symbol $s\supk_{\text{A}}$ overlap with symbol $s\supk_{\text{B}}$.
The contribution of user B to the $k$-th window of $N$ samples ${y}^{(k)}[n]$ can therefore be split into two parts,
namely $y\supk_{\text{B},1}[n]$ for $n \in \mathcal{N}_1 = \{0, \dots, \ctau - 1\}$ and
$y\supk_{\text{B},2}[n]$ for $n \in \mathcal{N}_2 = \{\ctau, \dots, N-1\}$ \cite{afisiadis2020advantage},
with
\begin{align}
        y\supk_{\text{B},1}[n] &= 
        e^{j 2 \pi\left( \frac{(n + N - \tau)^2}{2N} + (n + N - \tau) \left(\frac{s\supkM_{\text{B}}}{N} - \frac{1}{2} - u \left[ n-n\supk_{f,1} \right] \right) \right)}, \\
        y\supk_{\text{B},2}[n] &= 
        e^{j 2 \pi\left( \frac{(n + N - \tau)^2}{2N} + (n + N - \tau) \left(\frac{s\supk_{\text{B}}}{N} - \frac{1}{2} - u \left[ n-n\supk_{f,2} \right] \right) \right)}.
\end{align}

Prior to synchronization, both users are affected by distinct carrier frequency offsets relative to the receiver, namely $\Delta f_{c,\text{A}}$ and $\Delta f_{c,\text{B}}$.
However, since we assume the receiver to be synchronized to user A, there is a single effective CFO $\Delta f_{c} = \Delta f_{c,\text{B}} - \Delta f_{c,\text{A}}$ that only affects the signal from user~B.

Finally, the users have different transmit powers and experience independent channels $h_{\text{A}}$ and $h_{\text{B}}$. 
Thus, the baseband-equivalent model of the sampled signal contained in the $k$-th window of $N$ samples is
\begin{equation} \label{eq:signal-model}
    y\supk[n] = h_{\text{A}} x_{s\supk_{\text{A}}}[n] + z[n] +
    \left\{\begin{array}{ll}
        {h_{\text{B}} c[n] y\supk_{\text{B},1}[n],} & {n \in \mathcal{N}_1,} \\
        {h_{\text{B}}c[n] y\supk_{\text{B},2}[n],} & {n \in \mathcal{N}_2,}
    \end{array} \right.
\end{equation}
where $c[n] = e^{j 2 \pi n \frac{\Delta f_c}{f_s}}$ is the effective CFO affecting user B.

\section{A practical ML-derived multi-user receiver}
\label{sec:receiver}

\begin{figure}[t]
    \begin{tikzpicture}[shorten >=1pt,node distance=3.4cm,on grid,auto] 
        \node[state,minimum size=1.8cm] (q_0)   {No user}; 
        \node[state,minimum size=1.8cm] (q_1) [right=of q_0] {Single user}; 
        \node[state,minimum size=1.8cm] (q_2) [right=of q_1] {Two users}; 
         \path[->] 
         (q_0) edge[bend left=20] node [text width=1.5cm]{\footnotesize{New user: \color{red}\textit{synchronize}}} (q_1)
         (q_1) edge[bend left=20] node {\footnotesize{User leaves}} (q_0)
               edge[bend left=20] node {\footnotesize{New weak user}} (q_2)
               edge[bend left=60] node [text width=2.3cm]{\footnotesize{New strong user: \color{red}\textit{re-synchronize}}} (q_2)
         (q_2) edge[bend left=20] node {\footnotesize{Weak user leaves}} (q_1)
               edge[bend left=60] node [text width=2.3cm]{\footnotesize{Strong user leaves: \color{red}\textit{re-synchronize}}} (q_1);
     \end{tikzpicture}
     \caption{Finite state machine representation of the receiver. The receiver always synchronizes to the strongest user.}
     \label{fig:fsm}
\end{figure}
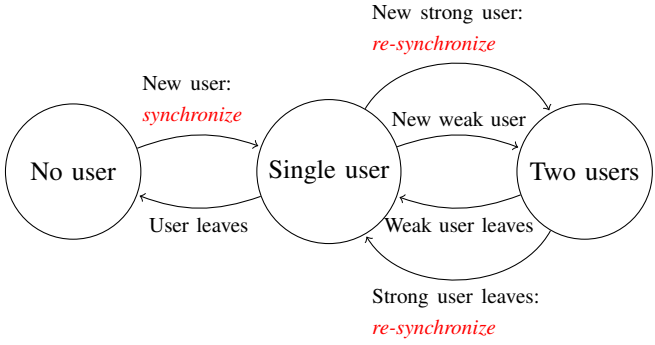

As previously explained, a LoRa user can start or stop transmitting at any time during the transmission of another user.
The gateway must hence be capable of tracking the arrival and departure of successive users. To this end, the proposed receiver implements a three-state finite state machine (FSM)
which is illustrated in Fig.~\ref{fig:fsm}. Each state corresponds to the current number of colliding users, up to two. 

In the no-user or single-user states, a synchronization algorithm is constantly running to detect the preamble of a potential arriving user.
When a new user is detected, the algorithm estimates its parameters, i.e., its CFO, STO, and power.
For the first user, the receiver synchronizes in time and frequency to this user, and the non-coherent detector of \eqref{eq:su-nc} is used to demodulate its symbols.
Upon the arrival of a second user, the receiver switches to a two-user detection rule capable of jointly demodulating both users.
Our two-user detector requires the receiver to re-synchronize to the strongest user.
It also requires the parameter estimation of the second user in the presence of interference.
In the following, we first explain the parameter estimation and synchronization stages and then discuss the two-user detection.

\subsection{Robust Synchronization in the Presence of Interference}

Contrary to the algorithms of \cite{xhonneux2019low} and \cite{tapparel2020open}, the synchronization stage of our two-user
receiver has to estimate the parameters of a new user even in the presence of a colliding user.
As explained in \cite{bernier2020low}, the estimation of the integer offsets $L_{\text{CFO}}$ and $L_{\text{STO}}$ requires
the demodulation of an upchirp and a downchirp in the preamble. 
However, the conventional single-user detector is typically not able to correctly demodulate these symbols in the presence of strong interference.

Hence, to perform a robust estimation of the CFO and STO, we propose instead to leverage the repetition of the upchirps
and the almost orthogonal relationship between the 2.25 downchirps in the preamble of the second user and the modulated symbols of the first user \cite{xu2020fliplora}.
The proposed algorithm implements the following steps:
\begin{enumerate}
    \item Instead of demodulating each upchirp of the preamble separately, the magnitudes of the DFTs of $N_{\text{pr}} - 1$
    consecutive windows of $N$ samples are averaged with a geometric mean. 
    \item The largest bin of the averaged DFTs is selected as demodulated upchirp. The algorithm uses this demodulated value to estimate $\lambda_{\text{CFO}}$
    over the $N_{\text{pr}} - 1$ upchirps with the estimator from \cite{bernier2020low}. The fractional CFO is corrected for the subsequent steps.
    \item The demodulation of the two downchirps is performed with a cross-correlation of length $2NR$ on an oversampled version of the
    received signal, where $R$ is the oversampling factor. The polyphase that yields the largest output is used as an estimate of $\lambda_{\text{STO}}$.
    The magnitude of the largest output is used to estimate the received power of the user (including the channel).
    \item The demodulated aggregated upchirp and the downchirps are used to estimate $L_{\text{CFO}}$ and $L_{\text{STO}}$, as described in \cite{bernier2020low}.
\end{enumerate}
If the second detected user is the strongest one, the receiver re-synchronizes
to this user by selecting the closest polyphase of the oversampled signal and correcting the estimated CFO.

\subsection{Maximum-Likelihood-based Two-User Detector}

We now present the detector used to jointly demodulate the symbols from two superimposed LoRa users.
The proposed detection rule is derived from the maximum likelihood sequence detector for the signal model given by \eqref{eq:signal-model}.

Without loss of generality, we assume that both users send a frame of $M$ symbols, and that the receiver is synchronized to user A.
This two-user $M$-symbol asynchronous signal can be decomposed into $M$ successive windows of $N$ samples,
and a final window of $\ctau$ samples.
Since every symbol from user B is contained in two consecutive windows,
the joint maximum likelihood decision of all symbols $\{ s^{(k)}_{\text{A}}, s^{(k)}_{\text{B}} \}$ for $k \in \{0, \dots, M-1\}$ implies to process all $M + 1$ windows at once.
The maximum likelihood sequence can be computed with reduced complexity using the Viterbi algorithm \cite{verdu1998multiuser}.
However, since there are $N^2$ possible states and $N^3$ possible state transitions, 
the complexity of this maximum likelihood sequence detector is still prohibitive, especially for large SFs (i.e., large $N$). 

To avoid such a complex receiver, we suggest to move from a joint demodulation of all symbols to individual decisions bounded to a single window of $N$ samples.
For each window $k$ and the corresponding received signal $y\supk$, an individual decision amounts to detecting the overlapping symbols $\boldsymbol{s}\supk = \{s\supk_{\text{A}}, s\supkM_{\text{B}}, s\supk_{\text{B}} \}$.
Let $P_{\text{A}}$ and $P_{\text{B}}$ represent the powers of user A and user B at the receiver, respectively.
Similarly, we define $\theta^{(k)}_{\text{A}}$ and $\theta^{(k)}_{\text{B}}$ as the initial phase of the symbols $s^{(k)}_{\text{A}}$ and $s^{(k)}_{\text{B}}$ at the receiver.
It can be shown that\footnote{We skip the detailed derivation due to space constraints.} the individual ML detection for a single window is achieved by selecting the symbols $\boldsymbol{\hat{s}}\supk $ that maximize the following metric 
\begin{equation} \label{eq:coherent-receiver}
\begin{aligned}
    \Lambda_{\text{ML}}(\boldsymbol{\bar{s}}\supk) &= \exp \Big[ \sqrt{P_{\text{A}}} \Re \Big( e^{-j \theta\supk_{\text{A}}} Y^{(k)} \left[ \bar{s}^{(k)}_{\text{A}} \right] \Big) \\
    &+ \sqrt{P_{\text{B}}} \Re \Big( e^{-j \theta\supkM_{\text{B}}} M^{(k)}_1 \left[ \bar{s}^{(k)}_{\text{A}}, \bar{s}^{(k-1)}_{\text{B}} \right] \Big) \\
    &+ \sqrt{P{_\text{B}}} \Re \Big( e^{-j \theta\supk_{\text{B}}} M^{(k)}_2 \left[ \bar{s}^{(k)}_{\text{A}}, \bar{s}^{(k)}_{\text{B}} \right] \Big) \Big],
\end{aligned}
\end{equation}
where $\boldsymbol{\bar{s}}\supk = \{\bar{s}\supk_{\text{A}}, \bar{s}\supkM_{\text{B}}, \bar{s}\supk_{\text{B}} \}$ are the candidate symbols evaluated by the ML criterion,
and $Y^{(k)}$, $M^{(k)}_1$ and $M^{(k)}_2$ correspond to three matched filters that are connected to each of the three symbols.
It is worth noting that this detector requires knowledge of the offsets $\tau$ and $\Delta f_c$, the phases $\{ \theta^{(k)}_{\text{A}}, \theta^{(k)}_{\text{B}} \}$
and the received powers of each user.

Since the receiver is synchronized to user A, the symbol $s^{(k)}_{\text{A}}$ translates to a Kronecker delta in the frequency domain.
The matched filter $Y^{(k)}$ hence corresponds to the DFT of the dechirped signal $\tilde{y}^{(k)}[n]$.
The two remaining terms in \eqref{eq:coherent-receiver} evaluate the contribution from user B, which is not synchronized.
As the symbols of this user lie partially in a different signal space than $s\supk_{\text{A}}$, the individual ML detector uses two specific matched filters $M^{(k)}_1$ and $M^{(k)}_2$
which depend on the offsets $\Delta f_c$ and $\tau$.
For both matched filters, we first eliminate the presumed contribution $\sqrt{P_{\text{A}}} e^{j \theta\supk_{\text{A}}} e^{j 2 \pi \frac{n}{N} \bar{s}^{(k)}_{\text{A}}}$ of user A,
and then compute a partial DFT over the first $\ctau$ and the remaining $N - \ctau$ samples in the window, respectively
\begin{align}
    M^{(k)}_1 \left[ \bar{s}^{(k)}_{\text{A}}, \bar{s}^{(k-1)}_{\text{B}} \right] & = \sum_{n=0}^{\ctau - 1} 
    \left( \tilde{y}^{(k)}[n] - \sqrt{P_{\text{A}}} e^{j \theta\supk_{\text{A}}} e^{j 2 \pi \frac{n}{N} \bar{s}^{(k)}_{\text{A}}} \right) \nonumber \\ 
    & \cdot e^{-j 2\pi \frac{n}{N} \left(\bar{s}^{(k-1)}_{\text{B}} - \tau + N\frac{\Delta f_c}{f_s} \right)} e^{j 2 \pi \tau u\left[n-n\supk_{f,1}\right]}, \\
    M^{(k)}_2 \left[ \bar{s}^{(k)}_{\text{A}}, \bar{s}^{(k)}_{\text{B}} \right] &=  \sum_{n=\ctau}^{N - 1}
    \left( \tilde{y}^{(k)}[n] - \sqrt{P_{\text{A}}} e^{j \theta\supk_{\text{A}}} e^{j 2 \pi \frac{n}{N} \bar{s}^{(k)}_{\text{A}}} \right) \nonumber \\ 
    & \cdot e^{-j 2\pi \frac{n}{N} \left(\bar{s}^{(k)}_{\text{B}} - \tau + N\frac{\Delta f_c}{f_s}\right)} e^{j 2 \pi \tau u\left[n-n\supk_{f,2}\right]}.
\end{align}

Yet, the matched filters $M^{(k)}_1$ and $M^{(k)}_2$ provide only partial information on the symbols $s^{(k-1)}_{\text{B}}$ and $s^{(k)}_{\text{B}}$,
as illustrated in Fig.~\ref{fig:interferences}.
Detecting these symbols using only the information from the $k$-th window is clearly suboptimal.
To efficiently demodulate the symbols $s^{(k-1)}_{\text{B}}$ and $s^{(k)}_{\text{B}}$, the detector must also use the partial information provided by 
the matched filters $M^{(k-1)}_2$ and $M^{(k+1)}_1$ of the preceding and following windows, respectively.
We hence suggest to 
estimate $s\supk_{\text{A}}$ by marginalizing over all candidates for $s\supkM_{\text{B}}$ and $s\supk_{\text{B}}$ independent of the previous and next window.
We then decide on $s\supkM_{\text{B}}$ based on the prior decisions on $s\supkM_{\text{A}}$ and on the latest decision on $s\supk_{\text{A}}$ as follows: 
\begin{align} \label{eq:demod-b}
    \hat{s}^{(k)}_{\text{A}} &= \max_{\bar{s}^{(k)}_{\text{A}}} \; \max_{\bar{s}^{(k-1)}_{\text{B}}, \bar{s}^{(k)}_{\text{B}}} \Lambda_{\text{ML}}(\bar{s}^{(k)}_{\text{A}}, \bar{s}^{(k-1)}_{\text{B}}, \bar{s}^{(k)}_{\text{B}}), \\
    \hat{s}^{(k-1)}_{\text{B}} &= \max_{\bar{s}^{(k-1)}_{\text{B}}}
    \left| M^{\scriptscriptstyle(k-1)}_2 \left[ \hat{s}^{\scriptscriptstyle(k-1)}_{\text{A}}, \bar{s}^{\scriptscriptstyle(k-1)}_{\text{B}} \right] + 
    M^{\scriptscriptstyle(k)}_1 \left[ \hat{s}^{\scriptscriptstyle(k)}_{\text{A}}, \bar{s}^{\scriptscriptstyle(k-1)}_{\text{B}} \right]  \right| .
\end{align}
The decision on $s\supk_{\text{B}}$ is deferred to the next time step.
This rule requires that the receiver keeps in memory the vector containing the $N$ matched filter outputs $M^{\scriptscriptstyle(k-1)}_2 \left[ \hat{s}^{\scriptscriptstyle(k-1)}_{\text{A}}, \bar{s}^{\scriptscriptstyle(k-1)}_{\text{B}} \right]$
of the previous window.

The individual ML detection rule given in \eqref{eq:coherent-receiver} requires the knowledge of the phases $\theta^{(k)}_{\text{A}}$, $\theta^{(k-1)}_{\text{B}}$
and $\theta^{(k)}_{\text{B}}$.
In practical systems, a residual CFO or a drifting STO often arise and modify the initial phases of the symbol during the transmission of the packet.
To avoid a continuous tracking of the phase of each user, we further propose to marginalize over all three phase terms in \eqref{eq:coherent-receiver}.
The marginalization of $\theta^{(k)}_{\text{A}}$ is however challenging, as it is used in all three parts of $\Lambda_{\text{ML}}(\boldsymbol{\bar{s}}\supk)$.
In a single-user scenario, the phase $\theta$ of a demodulated symbol $\hat{s}$ can be estimated 
by using the phase of the DFT bin $\hat{s}$, i.e., $\hat{\theta} = \arctan \left( Y\left[ \hat{s} \right]\right)$ \cite{afisiadis2019error}.
We propose to use the same estimator with the DFT $Y \supk$ to obtain an estimate $\hat{\theta}\supk_{\text{A}}$ of the initial phase of the candidate symbol $\bar{s}^{(k)}_{\text{A}}$.
Let $\widetilde{\Lambda}(\boldsymbol{\bar{s}}\supk)$ be a modified version of the function $\Lambda_{\text{ML}}(\boldsymbol{\bar{s}}\supk)$
where the variable $\theta^{(k)}_{\text{A}}$ is replaced by the estimate $\hat{\theta}\supk_{\text{A}}$ in the expression of the matched filters $M_1\supk$ and $M_2\supk$.
The remaining occurrence of $\theta^{(k)}_{\text{A}}$ in $\widetilde{\Lambda}(\boldsymbol{\bar{s}}\supk)$ is marginalized along with $\theta^{(k-1)}_{\text{B}}$ and $\theta^{(k-1)}_{\text{B}}$, yielding
\begin{equation} \label{eq:mu-detection}
\begin{aligned}
    \Lambda(\boldsymbol{\bar{s}}\supk) &= \iiint_{-\pi}^{\pi} \widetilde{\Lambda}(\boldsymbol{\bar{s}}\supk) \; \, d\theta\supk_{\text{A}} \, d\theta\supkM_{\text{B}} \, d\theta\supk_{\text{B}}\\
    & = I_0\left( \sqrt{P_{\text{A}}} \left| Y^{(k)} \left( \bar{s}^{(k)}_{\text{A}} \right) \right| \right) \cdot \\
    & I_0\left(\sqrt{P_{\text{B}}} \left| M^{(k)}_1 \big( \bar{s}^{(k)}_{\text{A}}, \bar{s}^{(k-1)}_{\text{B}} \big) \right| \right) \cdot \\
    & I_0\left( \sqrt{P{_\text{B}}} \left| M^{(k)}_2 \big( \bar{s}^{(k)}_{\text{A}}, \bar{s}^{(k)}_{\text{B}} \big) \right| \right),
\end{aligned}
\end{equation}
where $I_0(x)$ is the first order modified Bessel function of the first kind. This function is akin to $e^x$, and has a practical closed-form expression.
The accuracy of the estimate $\hat{\theta}\supk_{\text{A}}$ is increased when $P_A > P_B$, i.e., when the synchronized user is the strongest user.
This behavior is the principal motivation for the receiver to synchronize to the strongest user in the two-user state.

\section{Software-Defined Radio Implementation}
\label{sec:sdr}

The two-user detector and the synchronization algorithm previously described have been implemented
on the GNU Radio software-defined radio platform. We use this implementation in a testbed to experimentally assess the error rate performance
of the proposed receiver.
The implementation of our multi-user receiver is open-source and available at~\cite{2020GNUrepo}.

\subsection{Architecture of the Implementation}

The implementation of the two-user receiver is split into three stages.
The first stage performs the preamble detection and parameter estimation of new users on an oversampled signal, following the algorithm described in Section~\ref{sec:receiver}.
Upon detection of a new user, the preamble detection stage provides the estimated CFO, STO, and received power of the user to the synchronization stage.
The synchronization stage implements the FSM of Fig.~\ref{fig:fsm}. This stage stores in memory the parameters of the current users and synchronizes the receiver to the strongest user.
The synchronized signal is split into windows of $N$ samples, and each window is fed to the demodulation stage.
In the presence of two users, the receiver jointly demodulates both users using the metric $\Lambda(\boldsymbol{\bar{s}}\supk)$ of \eqref{eq:mu-detection},
and otherwise uses the non-coherent single-user detector of \eqref{eq:su-nc}.

\subsection{Testbed Description}

The testbed uses three National Instruments (NI) 2920 USRP transceivers, with two acting as transmitters and the third one implementing the two-user receiver.
The transmission of the LoRa frames is implemented using the open-source GNU Radio LoRa prototype described in \cite{tapparel2020open}, which has been tested to be compatible with commercial LoRa devices.
\begin{figure}[t] 
    \centering
    \includegraphics[width=0.4\textwidth]{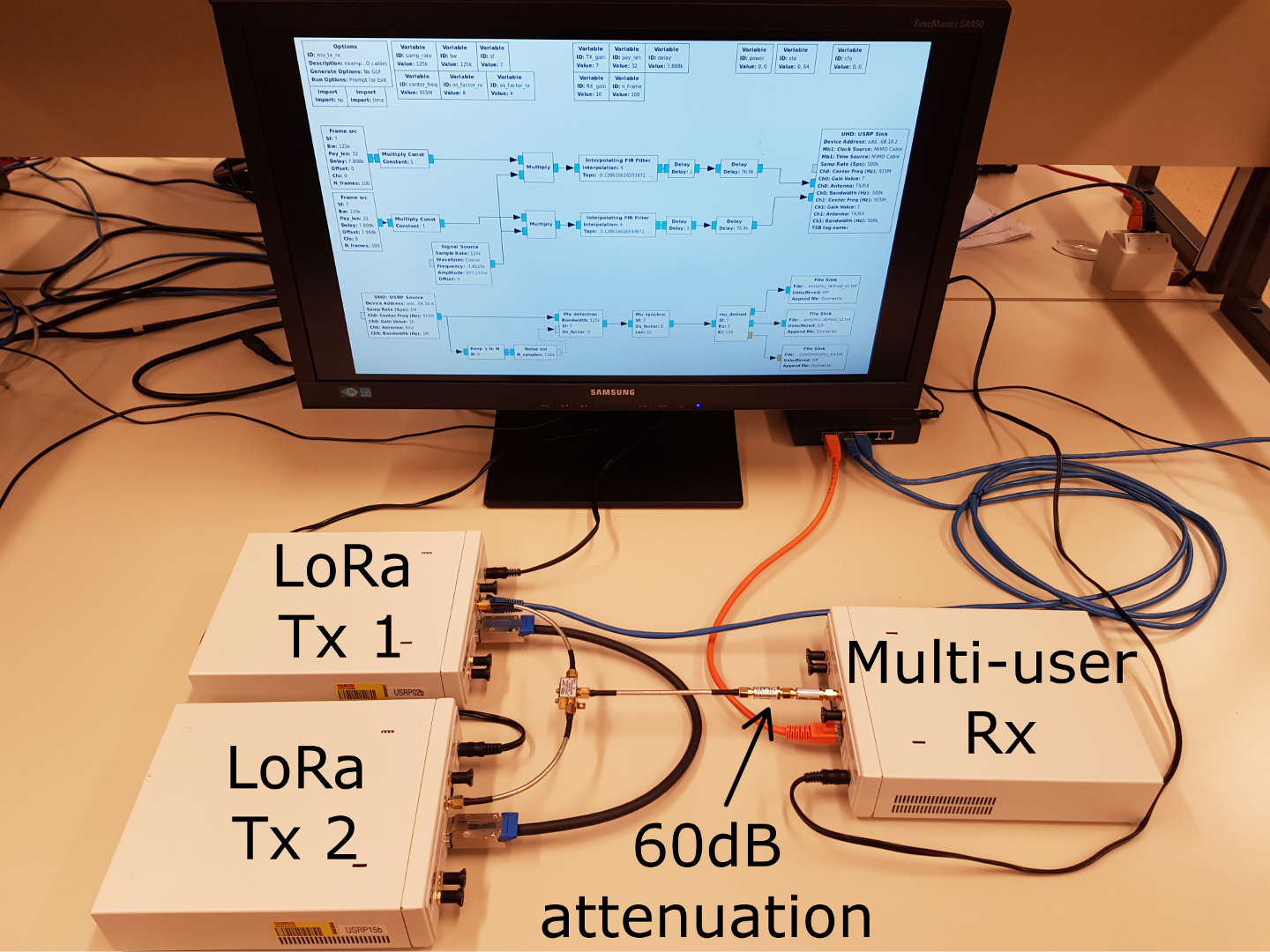}
    \caption{Multi-user testbed with two USRPs as Tx and one USRP as Rx.}
    \label{fig:testbed}
\end{figure}
The nominal carrier frequency used by both transmitters is 868MHz.  
To avoid interference from other sources in the 868MHz band, the transmitters are connected to the multi-user receiver with an RF combiner.
Two $-30$~dB attenuators are inserted between the combiner and the receiver to attain the low SNR regions of interest.
Both transmit USRPs share the same reference clock, which enables GNU Radio to define the STO $\tau$ and the effective CFO $\Delta f_c$ between the transmitters.
The CFO and STO between the transmitters and the receiver are not controlled in the testbed, but are corrected
by the receiver when it synchronizes to the strongest transmitter.

\section{Performance evaluation}
\label{sec:performance}

In the following, we provide measurements to evaluate the performance of the GNU Radio multi-user receiver using the testbed described in Section~\ref{sec:sdr}.
The experiments yield per-user symbol error rates (SER) when the receiver jointly demodulates two superimposed users.
While many parameters influence the SERs, in this paper we only focus on the impact of the STO $\tau$ between the users due to space constraints.

The results are obtained by creating collisions between the two transmit USRPs.
In each experiment, both transmitters send each one LoRa packet of $N_P = 32$ random symbols to the receiver, with $\textrm{SF} = 7$.
The starting time of the transmission of the second user is delayed by $15$ symbols and $\tau$ samples with respect to the first user.
The transmit power of the second user is always $3$~dB stronger than the power of the first user. The effective CFO $\Delta f_c$ between the transmitters is set to zero.
The receiver samples the received signal with an oversampling factor $R = 8$.
The SER vs SNR curves are obtained by sweeping through the transmit gains of the transmit USRPs. A total of $20\,000$ experiments are performed for each SNR level.
An experiment is considered to be valid if both packets are detected by the receiver and the estimated power of the second
user exceeds the estimated power of the first user. Invalid experiments are not taken into account in the SER evaluation.
Finally, only the $15$ overlapping payload symbols are used for the computation of the SERs.

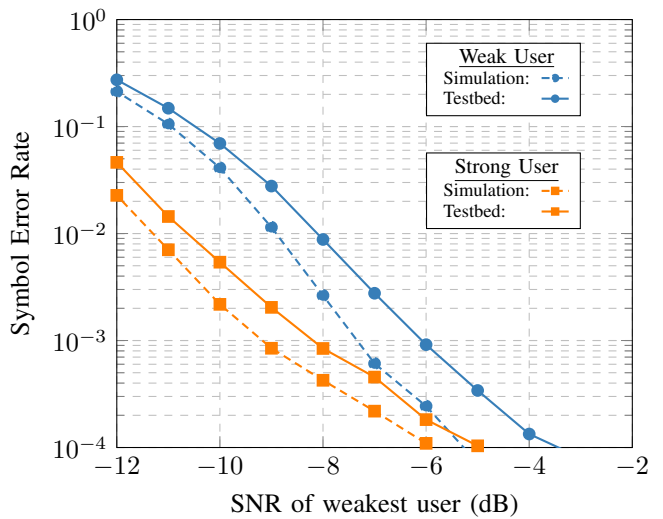
\begin{figure}[t] 
    \centering
    \begin{tikzpicture}

	\pgfplotsset{grid style={dashed}}

    \begin{semilogyaxis}[
        xlabel = {SNR of weakest user (dB)},
        ylabel = {Symbol Error Rate},
        ylabel near ticks,
        xlabel near ticks,
        xmin = -12, xmax = -2,
        ymin = 1e-4, ymax = 1,
        grid = both,
        legend image post style={scale=0.6}
        ]

		
        \addplot[Set1-7-2, thick, densely dashed, mark=*, mark options={scale=1}] table [x=SNR, y=SERu, col sep=comma] {results/theo_640.csv};
        \label{tau64_u1_MC}
        \addplot[Set1-7-5, thick, densely dashed, mark=square*, mark options={scale=1, solid}] table [x=SNR, y=SERi, col sep=comma] {results/theo_640.csv};
        \label{tau64_u2_MC}

        \addplot[Set1-7-2, thick, solid, mark=*, mark options={scale=1}] table [x=SNR, y=SERu, col sep=comma] {results/exp_640.csv};
		\label{tau64_u1_TB}
        \addplot[Set1-7-5, thick, solid, mark=square*, mark options={scale=1, solid}] table [x=SNR, y=SERi, col sep=comma] {results/exp_640.csv};
		\label{tau64_u2_TB}

		\node [draw,fill=white,inner sep=2pt] at (rel axis cs: 0.75,0.86) {
		\scriptsize
		\setlength{\tabcolsep}{2pt}
		\begin{tabular}{lc}
			\multicolumn{2}{c}{\footnotesize\underline{Weak User}} \\		
			Simulation:	& \ref{tau64_u1_MC} \\
			Testbed: & \ref{tau64_u1_TB}
        \end{tabular}
        };
		
		\node [draw,fill=white,inner sep=2pt] at (rel axis cs: 0.75,0.6) {
		\scriptsize
		\setlength{\tabcolsep}{2pt}
		\begin{tabular}{lc}
			\multicolumn{2}{c}{\footnotesize\underline{Strong User}} \\		
			Simulation:	& \ref{tau64_u2_MC} \\
			Testbed: & \ref{tau64_u2_TB}
		\end{tabular}};
	\end{semilogyaxis}

\end{tikzpicture}%
    \caption{Simulation and experimental SERs of two users with $\tau = 64.0$, $\Delta f_c = 0$, $P_{\text{A}} - P_{\text{B}} = 3$dB and $\textrm{SF} = 7$.}
    \label{fig:res64}
\end{figure}

Fig.~\ref{fig:res64} shows the experimental SERs of both users along with Monte-Carlo simulation results
for an STO between the users of $\tau = 64.0$ samples.
The simulation assumes perfect synchronization, i.e., the received power, CFO, and STO of each user are known.
The SER differences between the simulation and experimental results mainly illustrate the non-idealities of the synchronization stage.
We observe that for both users and in all SNR regimes, there is a loss of approximately $1$~dB between the simulation and the SDR implementation.
This result indicates that the synchronization algorithm is capable of estimating the parameters of both users.

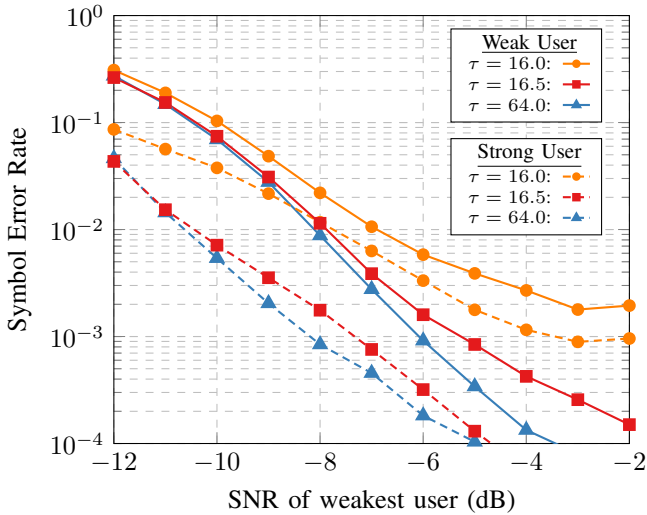
\begin{figure}[t]
    \centering
    \begin{tikzpicture}

	\pgfplotsset{grid style={dashed}}

    \begin{semilogyaxis}[
        xlabel = {SNR of weakest user (dB)},
        ylabel = {Symbol Error Rate},
        ylabel near ticks,
        xlabel near ticks,
        xmin = -12, xmax = -2,
        ymin = 1e-4, ymax = 1,
        grid = both,
        legend image post style={scale=0.6}
        ]

		
        \addplot[Set1-7-2, thick, solid, mark=triangle*, mark options={scale=1.4}] table [x=SNR, y=SERu, col sep=comma] {results/exp_640.csv};
        \label{tau640_u1}
        \addplot[Set1-7-2, thick, densely dashed, mark=triangle*, mark options={scale=1.4, solid}] table [x=SNR, y=SERi, col sep=comma] {results/exp_640.csv};
        \label{tau640_u2}

        \addplot[Set1-7-5, thick, solid, mark=*, mark options={scale=1}] table [x=SNR, y=SERu, col sep=comma] {results/exp_160.csv};
        \label{tau160_u1}
        \addplot[Set1-7-5, thick, densely dashed, mark=*, mark options={scale=1, solid}] table [x=SNR, y=SERi, col sep=comma] {results/exp_160.csv};
        \label{tau160_u2}

        \addplot[Set1-7-1, thick, solid, mark=square*, mark options={scale=1}] table [x=SNR, y=SERu, col sep=comma] {results/exp_165.csv};
        \label{tau165_u1}
        \addplot[Set1-7-1, thick, densely dashed, mark=square*, mark options={scale=1, solid}] table [x=SNR, y=SERi, col sep=comma] {results/exp_165.csv};
        \label{tau165_u2}

		\node [draw,fill=white,inner sep=2pt] at (rel axis cs: 0.80,0.86) {
        \scriptsize
		\setlength{\tabcolsep}{2pt}
		\begin{tabular}{lc}
			\multicolumn{2}{c}{\footnotesize\underline{Weak User}} \\		
			$\tau = 16.0$:	& \ref{tau160_u1} \\
            $\tau = 16.5$:	& \ref{tau165_u1} \\
            $\tau = 64.0$:	& \ref{tau640_u1}
        \end{tabular}
        };
		
		\node [draw,fill=white,inner sep=2pt] at (rel axis cs: 0.80,0.60) {
        \scriptsize
		\setlength{\tabcolsep}{2pt}
		\begin{tabular}{lc}
			\multicolumn{2}{c}{\footnotesize\underline{Strong User}} \\		
			$\tau = 16.0$:	& \ref{tau160_u2} \\
            $\tau = 16.5$:	& \ref{tau165_u2} \\
            $\tau = 64.0$:	& \ref{tau640_u2}
        \end{tabular}
        };
	\end{semilogyaxis}

\end{tikzpicture}%
    \caption{Experimental SERs of two users for different values of $\tau$, with $\Delta f_c = 0$, $P_{\text{A}} - P_{\text{B}} = 3$dB and $\textrm{SF} = 7$.}
    \label{fig:res16}
\end{figure}

We subsequently compare three scenarios with $\tau = 16.0$, $\tau = 16.5$ and $\tau = 64.0$ to study the influence of the STO on the demodulation. The
experimental SERs in all three scenarios are shown in Fig \ref{fig:res16}.
The case $\tau = 16.0$ is used as a baseline for the following discussion.
Regarding first the impact of the integer offset $L_{\text{STO}}$, we clearly observe that a larger integer STO (up to $\frac{N}{2}$) reduces the SERs of both users.
For the strongest user, the required SNR to attain a $10^{-3}$ SER, is $4$~dB lower for $\tau = 64.0$ compared to $\tau = 16.0$.
A similar behavior can be observed for the fractional part of the STO. With $\lambda_{\text{STO}} = 0.5$, the weakest user reaches a SER of $10^{-3}$
at $-5$~dB SNR, whereas for $\lambda_{\text{STO}} = 0$ 
the SER is much worse and even levels off an error floor around a SER of $10^{-3}$.

The observed behavior illustrates that the more the interfering users are desynchronized in time,
the easier it is for the two-user receiver to separate and demodulate them.
This effect can be explained by the contribution of each user to the DFT of the dechirped signal.
While the contribution of the strongest and synchronized user is always a Kronecker delta,
the signal space of the contribution of the weakest user depends on the STO $\tau$. For $\tau = 0$, both users share the same signal space and the matched
filter $M_1\supk$ is identical to the DFT $Y\supk$. In this case, it is difficult to distinguish the symbols of the two users.
In the presence of an STO $\tau \neq 0$, the contribution of the weakest user is no longer a single peak
but a bell-shaped function scattered across several DFT bins \cite{afisiadis2019error}. The maximum likelihood detector inherently leverages both parts of the STO
to separate the contribution of each user. Therefore, integer or fractional STOs close to $\frac{N}{2}$ or $0.5$, respectively, improve
the performance of the proposed two-user detector.

\section{Conclusion}

Multi-user receivers are required to overcome the scalability limitations of LoRa networks.
In this paper, we presented a receiver able to demodulate two colliding LoRa users.
This receiver is derived from the maximum likelihood two-user detector and relies on a novel synchronization algorithm robust to interference.
Instead of resorting to a costly maximum likelihood sequence estimation,
we propose a complexity reduction technique that enables the receiver to process one symbol at a time.
The proposed synchronization algorithm and two-user detector have been implemented on the GNU Radio SDR platform.
Experimental measurements show that the detector inherently leverages the time offset between the two interfering users
to separate and demodulate the contribution of each user.

\bibliographystyle{IEEEtran}
\bibliography{IEEEabrv,paper}

\end{document}